\documentclass[
sn-mathphys-num
]{sn-jnl}

\usepackage{graphicx}%
\usepackage{multirow}%
\usepackage{amsmath,amssymb,amsfonts}%
\usepackage{amsthm}%
\usepackage{tensor}
\usepackage{mathrsfs}%
\usepackage[title]{appendix}%
\usepackage{xcolor}%
\usepackage{textcomp}%
\usepackage{manyfoot}%
\usepackage{booktabs}%
\usepackage{algorithm}%
\usepackage{algorithmicx}%
\usepackage{algpseudocode}%
\usepackage{listings}%

\usepackage[utf8]{inputenc} 
\usepackage[english]{babel} 
\usepackage{makecell}
\usepackage{tabularx}

\usepackage{latexsym,bm}
\usepackage{amsmath,amsfonts}
\usepackage{amssymb}
\usepackage{tensor}
\usepackage{hyperref}

\usepackage{subfigure}
\usepackage[subfigure]{tocloft}
\usepackage{float}

\usepackage{dcolumn}

\usepackage{color}
\usepackage{tocloft}

\usepackage{amscd}							%
\usepackage{mathrsfs}						%
\usepackage{tikz-cd}						%
\usepackage[all,cmtip]{xy}					%
\usepackage{float}							%
\usepackage{mathtools}						%
\usepackage{bbm}							%
\usepackage[overload]{empheq}
\usepackage{lmodern}
\usepackage{thmtools}
\usepackage{enumitem}

\theoremstyle{thmstyleone}%
\newtheorem{example}{Example}%
\newtheorem{remark}{Remark}%
\newtheorem{theorem}{Theorem}
\newtheorem{lemma}[theorem]{Lemma}%
\newtheorem{definition}{Definition}%

\begin{document}

\title[Least Constraint and Contact Dynamics of Stochastic Vector Bundles]{Least Constraint and Contact Dynamics of Stochastic Vector Bundles}


\author*[1,2,3]{\fnm{D. Y.} \sur{Zhong}}\email{zhongdy@tsinghua.edu.cn}

\author[1]{\fnm{G. Q.} \sur{Wang}}\email{dhhwgq@tsinghua.edu.cn}


\affil*[1]{\orgdiv{State Key laboratory of Hydroscience and Engineering}, \orgname{Tsinghua University}, \orgaddress{\street{Qinghua Yuan Road}, \city{Haidian District}, \postcode{100084}, \state{Beijing}, \country{China}}}

\affil[2]{\orgdiv{Key Laboratory of Hydrosphere Sciences of the Ministry of Water Resources}, \orgname{Tsinghua University}, \orgaddress{\street{Qinghua Yuan Road}, \city{Haidian District}, \postcode{100084}, \state{Beijing}, \country{China}}}

\affil[3]{\orgdiv{Department of Hydraulic Engineering}, \orgname{Tsinghua University}, \orgaddress{\street{Qinghua Yuan Road}, \city{Haidian District}, \postcode{100084}, \state{Beijing}, \country{China}}}



\abstract{

This paper investigates the contact structures and dynamics of stochastic vector bundles, leading to the formulation of the least constraint theorem. It is found that the probability space of stochastic vector bundles possesses an infinite-order jet structure, which enables the geometric analysis of stochastic processes. Furthermore, this study demonstrates that stochastic vector bundles have a natural contact structure, leading to the decomposition of the tangent space and providing insight into the evolution and constraints of the system. Finally, we derive a set of contact dynamical equations for the stochastic vector bundles. These equations correspond to the least constraint on the evolution of stochastic vector bundles, which is a counterpart to the least action principle for symplectic structures. This shows the relationship between the geometric structure of the stochastic system evolution and its tendency to minimize constraints. This study provides a geometric framework for analyzing stochastic space with potential applications in various fields where probabilistic behavior is crucial.
}




\maketitle
\tableofcontents
%
%
%
\section{Introduction}\label{sec-1}

Stochastic processes are widely observed in multiple disciplines, including physics, biology, engineering, and finance. Studying stochastic processes can deepen our understanding of inherent uncertainties in real-world systems \cite{risken1984fokker, van1992stochastic}. It can also enable us to create more accurate models, develop better control strategies, and establish more effective risk management practices.

Studies on stochastic processes traditionally focus on analyzing their statistical properties, as they are essential for describing random phenomena that evolve over time or space \cite{Sto-m-cgispinGardiner}. Key statistical properties, such as the mean, variance, correlation, and higher-order moments, offer crucial insights into the behavior of these processes. Researchers have investigated probability distributions, expected values, and variances to predict future behavior, assess risks, and make informed decisions under uncertainty. The emphasis on statistical properties has established a robust foundation for understanding and modeling random phenomena, facilitating the development of practical tools for analyzing and predicting their behavior \cite{ zwanzig2001nonequilibrium}.

In addition to classical methods, other theories and techniques have been adopted to study stochastic processes. For example, differential geometry is one of them. Differential geometry has become ubiquitous across various scientific disciplines, from physics to computer science, owing to its powerful capacity to describe and analyze geometric structures and their transformations \cite{Nakahara2003GeometryTA, RUD2013, RUD2017}. This mathematical theory has also been successfully applied to the study of stochastic processes, leading to an increasing number of studies employing techniques developed in differential geometry to explore the intricacies of randomness and variability. Significant advances have been made in applying differential geometry to stochastic processes \cite{10.1007/978-94-010-2675-8_4, PINSKY1978199, Hsu-stochaticmanifold}. Some of these studies focused on stochastic processes on manifolds  
\cite{MIC1990, PINSKY1978199, Hsu-stochaticmanifold}, which locally resemble Euclidean space but have more complex global structures. Applying the principles of differential geometry to studies of stochastic processes allows us to understand the behavior of stochastic systems constrained by non-Euclidean geometries, thereby advancing our knowledge of complex phenomena in mechanics, thermodynamics, and biological systems, where configuration space is typically represented by various manifolds.

Nonetheless, the geometric structures of the probability space have received comparatively less attention. The fundamental concept of stochastic processes is the probability space, defined by the triplet \((\Omega, \mathcal{F}, \mathbb{P})\), where \(\Omega\) is the space of outcomes, \(\mathcal{F}\) is the \(\sigma\)-algebra of events, and \(\mathbb{P}\) is the space of probability measures. Exploring the geometric characteristics of the probability space at greater depths is essential because these properties can be vital when dealing with measures that do not conform to the standard Euclidean space. For example, in information geometry, which employs differential geometric techniques to examine probability distributions, the Fisher-Rao metric and its related geometric constructs help analyze the space of probability distributions \cite{WOS:000483732700021}. Moreover, the broader geometric aspects of probability spaces, particularly within the context of non-equilibrium stochastic processes of high-dimensional scenarios, are yet to be adequately investigated. Consequently, there is an increasing awareness of the necessity for additional research into the geometric structure of probability spaces, which promises to enrich ongoing studies of stochastic processes on manifolds.

Our previous study found an interesting connection between probability space and differential geometry, demonstrating that the probability space of a vector bundle inherently possesses the geometric structure of jet bundles \cite{Zhong_2024}. This study bridges the gap between abstract probability space and the underlying tangible geometry of vector space. The vector bundle, a topological construction generalizing the notion of a vector space, serves as a natural habitat for stochastic processes, where the probability space is not merely a collection of outcomes but is equipped with a series of rich geometric structures in terms of jet bundles. The jet bundle aspect of this probability space enables us to consider the points on the manifolds where the rates of change and higher-order derivatives of the probability density function are significant. 

Furthermore, as reported in \cite{Zhong_2024}, our study indicates that a stochastic vector bundle can possess a natural contact structure. For the vector space \(E\) and probability measure space \(\mathbb{P}\), the infinite-order jet bundle \(J^{\infty}(E, \mathbb{P})\) based on \(E\) for \(\mathbb{P}\) has the connection 1-form \(\Theta_{\wp}=\mathrm{d}P-\wp_i \mathrm{d}y^i\) defined on \(T^{*}J^{\infty}(E, \mathbb{P})\), where \(\wp\) is the local connection \cite{Zhong_2024}. The 1-form \(\Theta_{\wp}\) satisfies the condition that \(\Theta_{\wp} \wedge (\mathrm{d}\Theta_{\wp})^n\) is nowhere zero, which implies that the stochastic vector bundles have contact structures \cite{Geiges2008}. A contact manifold is a mathematical structure consisting of a \(2n+1\)-dimensional smooth manifold equipped with a smooth, maximally non-integrable vector field on the hyperplane of codimension-1 \cite{ArnoldVI1989, anold2010, Bravetti-CONTACTHAMILTONIAN, RivasXavier-JGE}. Contact geometry is closely related to symplectic geometry; however, several key differences exist \cite{anold2010}. In symplectic geometry, the focus is on the closed 2-form \(\omega\) of a 2n-dimensional manifold. In contrast, contact geometry deals with a 1-form \(\Theta\) and its exterior derivative \(\mathrm{d}\Theta\). The non-degeneracy condition in symplectic geometry corresponds to the maximally non-integrable condition in contact geometry.

This study was motivated by curiosity regarding how the probability space relates to contact geometry and whether it provides a new approach to understanding stochastic processes. Based on our previous studies \cite{Zhong_2024}, this study discusses the intricate structure of stochastic vector bundles, which can be leveraged to enhance our understanding and analysis of these systems. 

The remainder of this paper is organized as follows. In Section \ref{sec-2}, we introduce a jet bundle into the probability space of vector bundles. The jet bundle encapsulates the general geometric properties of stochastic vector bundles, which provides a theoretical framework for studying contact structures. In Section \ref{sec-3}, the contact structure of stochastic vector bundles is discussed. This fundamental geometric object decomposes the tangent space into two complementary subspaces: the contact distribution and its orthogonal complement. This decomposition is critical to understanding the evolution of a system and its constraints. In Section \ref{sec-4}, we derive a set of contact dynamical equations for the stochastic vector space that lies at the heart of contact dynamics. These equations resemble Hamilton's equations in form, but have significant differences: the Hamiltonian is replaced by a term representing the probability variation. At the end of the paper, in Section \ref{sec-5}, we provide the theorem of least constraint for stochastic vector bundles. This indicates a significant connection between the contact dynamical equations governing the stochastic vector bundles and the least constraint. This suggests that as the system evolves, it naturally favors trajectories that minimize constraints—i.e., residual energies—while maximizing the evolution of probability distributions. In Section \ref{sec-6}, we present the summaries and conclusions of this study. 

This study aims to provide a geometric framework for analyzing stochastic systems applicable across different fields, where probabilistic behavior plays a key role. Applications including direct areas, such as stochastic processes, generative models, transfer learning, anomaly detection, and robust reinforcement learning, are also possible targets.

\section{Stochastic Vector Bundles}\label{sec-2}
This section discusses the definition of a stochastic vector bundle and the kinetic equation governing its evolution. 

Stochastic processes, known for their unpredictability, are studied predominantly using statistical properties, such as moments, distributions, and other statistical parameters \cite{Sto-m-cgispinGardiner}. Non-equilibrium processes are mainly analyzed using kinetic equations such as the master equation and Fokker-Planck equation, with Langevin equations bridging deterministic and stochastic models 
\cite{risken1984fokker, van1992stochastic}. They are vital for understanding phenomena like fluctuation-dissipation and non-equilibrium phase transitions 
\cite{WOS:000073767900025, WOS:A1993LV74300001}, and for simulating complex systems 
\cite{WOS:000166192000001, Pope_2000, ZHONG-2022-KINETICEQAUTION}. The Fokker-Planck equation is a standard tool for non-equilibrium probability density functions, though it is limited to Markovian, Euclidean space processes 
\cite{risken1984fokker}. Recently, we extended the kinetic equation to stochastic vector bundles on general manifolds to examine non-Markovian, non-Gaussian, and non-stable stochastic processes \cite{Zhong_2024}. Our study showed that the stochastic vector bundle possesses rich geometrical properties inherent to jet bundles, as outlined below.
\\
\begin{definition}[Stochastic vector bundles]
Let \((\Omega, \mathcal{F}, \mathbb{P})\) denote the probability space with sample space \(\Omega\), \(\sigma-$algebra \(\mathcal{F}\), and probability measure space \(\mathbb{P}\). A stochastic process is an inclusive map from \((\Omega, \mathcal{F})\) by filtering \(\{ \mathcal{F} \}_{t\in T}\subseteq \mathcal{F}\) to a manifold \(M\) of dimension \(n\), that is, \(\gamma: \Omega \hookrightarrow M\). If a measurable vector \(Y: M \to E; \gamma\mapsto Y\circ\gamma \subset E\) is defined, \(\pi\circ Y=Id_M\) defines a stochastic vector bundle \(\pi: E\to M\). 
\\
\end{definition}
\begin{remark}
In this study, \(Y\) is assumed to be measurable, implying that it is assigned a probability measure \(P\in\mathbb{P}\) for \(Y\in U\subset E\), denoted by \(P(Y\in U)\equiv P\circ \gamma^{-1}\circ Y^{-1}(U)\).
\\
\end{remark}

Our study demonstrated that the probability measure \(\mathbb{P}\) of stochastic vector bundles has an infinite-order jet structure. Let \(\pi^{\infty}_{E,0}: J^{\infty}(E, \mathbb{P})\to E\) be a jet bundle of infinite-order. A point \(y \in E\) represents the realization of \(Y\); thus, \(P(Y(t)=y)\) is the probability that \(Y(t)\) coincides with \(y\). This is also understood as the realization of the stochastic variable \(Y\). Let \(j^{\infty}_yP=j^{\infty}P(y)\in J^{\infty}(E, \mathbb{P})\) denote the point of the jet bundle, then we have a series of local coordinate functions \((y^i,p, p_{\mu_1},p_{\mu_1\mu_2}, \cdots, p_{\mu_1\mu_2\cdots\mu_{\infty}})\) for \(j^{\infty}_yP\), by which \(j^{\infty}_yP\in J^{\infty}(E, \mathbb{P})\) has the local coordinates
\begin{equation}\label{eq-2-1}
(y^i,p, p_{\mu_1},p_{\mu_1\mu_2}\cdots,p_{\mu_1\mu_2\cdots\mu_{\infty}}): y\mapsto
\left\{
\begin{aligned}
	&y^{i}(j^{\infty}_yP)=y^i,\\
	&p(j^{\infty}_yP)=P(y),\\
	&p_{\mu_1}(j^{\infty}_yP)=\partial_{\mu_1}P(y),\\
		&\cdots\\
	&p_{\mu_1\mu_2\cdots\mu_{\infty}}(j^{\infty}_yP)=\partial_{\mu_1\cdots\mu_k}P(y),
\end{aligned}
	\right.
\end{equation}
where \(\partial_{\mu_k}\) stands for \(\tfrac{\partial}{\partial y^{\mu_k}}\), and \(	
	\partial_{\mu_1\cdots\mu_{\infty}}P(y)=\partial_{\mu_1}\otimes\cdots\otimes\partial_{\mu_{\infty}}P(y)\).

To establish the immediate relationships between \(E\) and \(J^{\infty}(E, \mathbb{P})\) without too many notions involved, in this paper, we define that 
\begin{equation}\label{eq-2-2}
	P_{\mu_1\cdots\mu_k}=p_{\mu_1\cdots\mu_k}\circ j^{\infty}\circ P, \ 0\le k \le\infty,
\end{equation}
which acts as a map \(P_{\mu_1\cdots\mu_{\infty}}: y\in E\mapsto \partial_{\mu_1\cdots\mu_{\infty}}P(y)\), and thus, we had a series of new local coordinate functions for \(\pi^{\infty}_{E,0}: J^{\infty}(E, \mathbb{P})\to E\), i.e., \((y^{i}, P, P_{\mu_1},\cdots, P_{\mu_1\cdots\mu_\infty})\), by which \(y\in  E\) is mapped to \(J^\infty(E, \mathbb{P})\) as
\begin{equation}\label{eq-2-3}
(y^{i}, P, P_{\mu_1},\cdots, P_{\mu_1\cdots\mu_\infty}):y \mapsto
	\left\{
	\begin{aligned}
		    &y^i(y)=y^i,\\
			&P(y)=P(y),\\
			&P_{\mu_1}(y)=\partial_{\mu_1}P(y),\\
			&\cdots\\
			&P_{\mu_1\cdots\mu_{\infty}}(y)=\partial_{\mu_1\cdots\mu_{\infty}} P(y).		
	\end{aligned}
	\right.
\end{equation}
The aforementioned jet bundle is illustrated using the commutative diagram shown in Figure \ref{Fig:Fig-1} \cite{Zhong_2024}.
\begin{figure}[h!]
\centering
\begin{tikzcd}[ column sep=8em, row sep=9em]
{}
&\mathbb{G} 
	\arrow[sloped]{ld}{j_y^{\infty}\circ R_P}
	\arrow[swap]{d}{R_P}
&\\
J_{E}^{\infty}\mathbb{P}
	\arrow[yshift=-0.5ex,swap]{r}{\pi^{\infty}_E}
&{\mathbb{P}}
	\arrow[yshift=0.5ex,swap]{l}{j_y^{\infty}}
	\arrow[yshift=-0.5ex,swap]{r}{\pi_E}
&E
	\arrow[yshift=0.5ex,swap]{l}{P\circ (Y \circ \gamma)^{-1}}
	\arrow[xshift=-0.5ex,swap]{d}{\pi}
	\arrow[sloped]{lu}{\mathrm{exp}(\int \pi^{*}\mathscr{L})}
	\\
&(\Omega, \mathcal{F})
	\arrow[hook]{r}{\gamma}
	\arrow[sloped]{ur}{Y\circ\gamma}
	\arrow{u}{P} 
	\arrow[sloped, swap]{lu}{j_y^{\infty}\circ P}
&M
	\arrow[xshift=0.5ex,swap]{u}{Y}
\end{tikzcd}
\caption{The commutative diagram for the jet bundle  \(\pi_{E, 0}^{\infty}: J_{E}^{\infty}\mathbb{P}\to E\), \(\pi_{E, 0}^{\infty}=\pi_E\circ\pi_E^{\infty}\). 
\(\gamma\) is a stochastic processes defined on \((\Omega, \mathcal{F})\); \(Y: M\to E\) is the section of \(\pi: E\to M\); \(P: (\Omega, \mathcal{F})\to \mathbb{P}\) is the probability measure; \(j_y^{\infty}\) is the jet of \(P\) that \(j_{y}^{\infty}: P\mapsto j_{y}^{\infty}P \in J_{E}^{\infty}\mathbb{P}\); \(R_P:  \mathbb{G}\to  \mathbb{P}\) is an inclusion; the mapping \(\exp(\int \pi^{*}\mathscr{L}): E\to \mathbb{G}\) constitutes a group \(\mathbb{G}\) acting on \(\mathbb{P}\).
}
\label{Fig:Fig-1}
 \end{figure}

Upon defining the stochastic vector bundles, we derived a kinetic equation governing the probability \(P\in \mathbb{P}\) \cite{Zhong_2024}. The equation takes the form of a partial differential equation of infinite order, as detailed below:
\begin{equation}\label{eq-3-1}
	\frac{\partial P(y,t)}{\partial t}=\mathscr{L}P(y,t),
\end{equation} 
where \(\mathscr{L}\) is the time-evolution operator defined as
\begin{equation}\label{eq-3-2}
	\mathscr{L}=-D^{\mu_1}\partial_{\mu_1}
	+\frac{1}{2!}D^{\mu_1\mu_2}\partial_{\mu_1\mu_2}+\cdots.
\end{equation}
Hereafter, the Einstein summation convention is used; that is, summation is performed over any index that appears twice in a single term. The coefficients \(D^{\mu_1\cdots\mu_k}\) (\(1\le k \le \infty\)) are the functionals of the cumulants concerning the state transition path \(S\) of the system, i.e., 
\begin{equation}\label{eq-3-3}
	D^{\mu_1\cdots\mu_n}=\frac{1}{n!}\frac{\partial}{\partial t}\langle\langle S^{\mu_1} \cdots S^{\mu_n}(\tau)\rangle\rangle,\quad
	S^{\mu}=
\int_0^t \mathcal{L}(y^{\mu}),
\end{equation}
where \(\mathcal{L}=\omega\otimes Y\) is a vector-valued 1-form, with \(Y\) representing the stochastic processes as a section of the vector bundle \(\pi: E\to M\) and \(\omega\) the connection 1-form. The details of the derivation of equations \eqref{eq-3-1} and \eqref{eq-3-3} are given in \cite{Zhong_2024}.

%
%
%
%
%
%

Our study also showed that the kinetic equation for stochastic vector bundle also determines the geodesic equation of $J^{\infty}(E, \mathbb{P})$ \cite{Zhong_2024}. It is to say that there exists a connection 1-form $\Theta_{\wp}\in T^{*}J^{\infty}(E, \mathbb{P})$ defined as
\begin{equation}\label{eq-3-4}
	\Theta_{\wp}
	=\mathrm{d}P_{\mu_1\cdots\mu_k}
	-\wp_i(y, P_{\mu_1\cdots\mu_k} )\mathrm{d}y^i, \  0\le k\le \infty,
\end{equation}
where \(\wp_i(y, P_{\mu_1\cdots\mu_k} )\) is the local connection and is usually assumed to be linearly dependent on \(P_{\mu_1\cdots\mu_k}\). For any $X\in TJ^{\infty}(E, \mathbb{P})$ given in the local coordinates \( (y^i, P_{\mu_1\cdots\mu_k}) \) as
\begin{equation}\label{eq-3-5}
\begin{split}
	X&=\dot{y}^i\frac{\partial }{
	\partial y^i}
	+\sum_{0\le k\le \infty}\sum_{\mu_1<\cdots<\mu_k}\dot{P}_{\mu_1\cdots\mu_k}\frac{\partial}{\partial P_{\mu_1\cdots\mu_k}}\\
	&=\dot{y}^i
	\left(
	\frac{\partial }{\partial y^i}
	+\sum_{0\le k\le \infty}\sum_{\mu_1<\cdots<\mu_k}P_{\mu_1\cdots\mu_ki}\frac{\partial}{\partial P_{\mu_1\cdots\mu_k}},
	\right),		
\end{split}
\end{equation}
the relation $\iota_{X}\Theta_{\wp}=0$ leads to 
\begin{equation}\label{eq-3-6}
	\dot{P}_{\mu_1\cdots\mu_k}
	=\frac{\mathrm{d}P_{\mu_1\cdots\mu_k}(t,y(s))}{\mathrm{d}s}\Big{|}_{s=t}
	=\wp_i(y, P_{\mu_1\cdots\mu_k})\mathrm{d}y^i, \ 0\le k\le \infty,
\end{equation}
where \(s\) is a parameter usually referred to as label time. For the case of \(k=0\), the connection 1-form is
\begin{equation}\label{eq-3-7}
	\Theta_{\wp}
	=\mathrm{d}P-\wp_i \mathrm{d}y^i,
\end{equation}
and $\iota_X\Theta_{\wp}=0$ leads to the geodesic equation of \(\mathbb{P}\) as
\begin{equation}\label{eq-3-8}
	\frac{\mathrm{d}P(t, y(s))}{\mathrm{d}s}
	=\wp_i\dot{y}^i.
\end{equation}
As shown in Appendix \ref{app-1}, the local connection \(\wp\in T^{*}J^{\infty}(E, \mathbb{P})\) as a 1-form is given by
\begin{equation}\label{eq-3-9}
	\wp=\wp_i\mathrm{d}y^i
	=\sum_{1\le k\le \infty}\sum_{\mu_1<\cdots<\mu_k} {B}^{\mu_1\cdots\mu_k}_iP_{\mu_1\cdots\mu_k}\mathrm{d}y^i.
\end{equation}
In the theory of fiber bundles, the connection 1-form defines a decomposition of \(TJ^{\infty}(E,\mathbb{P})\) at \(y\in E\) as a direct sum:
\begin{equation}\label{eq-3-10}
	TJ^{\infty}_y(E,\mathbb{P})=H \oplus V,
\end{equation}
of which, 
\begin{equation}\label{eq-3-11}
	H=\{X|\iota_X\Theta_{\wp}=0, X\in   TJ^{\infty}_y(E,\mathbb{P})\}.
\end{equation}

\section{Contact Structures}\label{sec-3}

This section discusses the contact structures of the stochastic vector bundles. 

Contact structures are critical geometrical objects in studies of dynamical systems \cite{ FOUNDATIONOFMECH, Bravetti-CONTACTHAMILTONIAN, RivasXavier-JGE, Geiges2008, anold2010}A contact structure on an oriented smooth manifold of odd dimension \(2n+1\) is a completely non-integrable hyperplane \(H\), which is a codimension-1 subbundle of the tangent bundle, such that the 1-form \(\Theta\) satisfies \( \Theta \wedge (d\Theta)^n \) is a nowhere vanishing volume form. This condition ensures that, at every point \(p\) of the manifold, \( H|_p = \ker(\Theta|_p) \) is a maximal distribution with \(\dim(H) = 2n\). The contact form satisfies the following conditions \cite{Bravetti-CONTACTHAMILTONIAN, Geiges2008, anold2010}: 

\begin{enumerate}
\item[(1)] \textbf{Non-degeneracy}. The contact form \(\Theta\) is non-degenerate in terms of the definition of the distribution, meaning that for each point \(p\), the 1-form \(\Theta|_p\) is applied to any vector \(X \in H|_p\) in the contact distribution \(H\) is zero if and only if \(X = 0\).

\item[(2)] \textbf{Volume form condition}. Contact form \(\Theta\) must satisfy the condition that \(\Theta \wedge (d\Theta)^n\) is a volume form on the manifold. This means that \(\Theta \wedge (d\Theta)^n\) is a non-vanishing \(2n+1\)-form, which is essential for the contact structure to be well-defined.

\item[(3)] \textbf{Maximal non-integrability}. The contact form is such that \(\mathrm{d}\Theta\) is non-vanishing when restricted to the contact distribution \(H\). This implies that the distribution cannot be spanned by the flows of a collection of functions (i.e., it is maximally non-integrable).
\end{enumerate}
\vspace{1em}
\begin{lemma}[Contact Structure on Infinite-Order Stochastic Jet Bundles]\label{lem-1}
Let \(E\) be a vector bundle over a manifold \(M\), and let \(\mathbb{P}\) denote the probability measure. For \(P\in \mathbb{P}\), the connection 1-form
\( \Theta_{\wp}=\mathrm{d}P-\wp_i\mathrm{d}y^i \) 
on the infinite-order stochastic jet bundle \(J^{\infty}(E, \mathbb{P})\) defines a natural contact structure.
\end{lemma}
\vspace{1em}
\begin{proof}
We start by noting that the connection 1-form \(\Theta_{\wp} = \mathrm{d}P - \wp_i \mathrm{d}y^i\) is defined on the infinite-order stochastic jet bundle \(J^{\infty}(E, \mathbb{P})\), where \(\wp = \wp_i \mathrm{d}y^i\) represents the local connection form. 

To show that \(\Theta_{\wp}\) defines a contact form, we apply Darboux's theorem \cite{Geiges2008}. This theorem implies that for a (2n+1)-dimensional manifold, which in this case is a submanifold of \(J^{\infty}(E,\mathbb{P})\), there exists a local coordinate system in which the contact 1-form takes a standard form \( \Theta_{\wp}=\mathrm{d}P-\wp_i\mathrm{d}y^i \). The contact structure is given by the distribution \(H = \{X \in TJ^{\infty}_y(E,\mathbb{P}) \mid \iota_X \Theta_{\wp} = 0\}\), which is equivalent to the kernel of \(\Theta_{\wp}\), i.e., \(\mathrm{ker}(\Theta_{\wp})\). To confirm that this distribution is a contact structure, we must verify that it is completely non-integrable and that the volume form induced by \(\Theta_{\wp}\) is non-degenerate.

First, we assume that \(P\) is non-constant, which ensures that \(\mathrm{d}P \neq 0\). Furthermore, it is trivial to find the \(\iota_{X}\Theta_{\wp}=0\), if and only if \(X=0\). This is a necessary condition for \(\Theta_{\wp}\) to be a contact form, as a contact form must be non-vanishing.

Second, we consider the non-degeneracy of the volume form. The condition \(\iota_X (\mathrm{d}\wp_i \wedge \mathrm{d}y^i) = 0\) for a vector field \(X\) implies that \(X\) must be zero, as \(\mathrm{d}\wp_i \wedge \mathrm{d}y^i\) is a non-degenerate 2-form. This non-degeneracy is crucial for the volume form induced by \(\Theta_{\wp}\) to be non-degenerate as well. To see this explicitly, we compute the volume form:
\[
\Theta_{\wp} \wedge (\mathrm{d}\Theta_{\wp})^n = (-1)^n \mathrm{d}P \wedge \mathrm{d}\wp_1 \wedge \mathrm{d}y^1 \wedge \cdots \wedge \mathrm{d}\wp_n \wedge \mathrm{d}y^n.
\]
Since \(\mathrm{d}P \neq 0\) and \(\mathrm{d}\wp_i \wedge \mathrm{d}y^i\) is non-degenerate, the above volume form is non-degenerate on \(TJ^{\infty}(E, \mathbb{P})\).

Third, Frobenius's theorem asserts that the non-degeneracy of the volume form establishes the complete non-integrability of the distribution \(H\). 

As shown above, we demonstrate that the connection 1-form \(\Theta_{\wp}\) indeed induces a natural contact structure on the infinite-order stochastic jet bundle \(J^{\infty}(E, \mathbb{P})\), as previously stated.
\\
\end{proof}

Further examination of the intricacies of the contact structure on \(J^{\infty}(E,\mathbb{P})\) is advisable, as it reveals the geometric foundation of phase space of stochastic vector bundles. Phase space, a fundamental concept in classical mechanics and various domains of physics, is a mathematical space where each point represents a unique state of a system defined by its position and momentum coordinates \cite{anold2010}. Understanding phase space is crucial because it elucidates how these coordinates interact. For stochastic vector bundles \(J^{\infty}(E, \mathbb{P})\), phase space has different meanings from that of classical mechanics. Investigating it will allow for a thorough encapsulation of stochastic properties, making it indispensable for analyzing and predicting the behavior of stochastic systems.

We started with a general 1-form \(\Theta_J \in T^{*}J^{\infty}(E,\mathbb{P})\) which will be illustrated as an isomorphism of \(\Theta_{\wp}\). It is defined as 
\begin{equation}\label{eq-3-12}
	\Theta_J=P_i\mathrm{d}y^i-
	\sum_{0\le k\le \infty}\sum_{\mu_1<\cdots<\mu_k} B^{\mu_1\cdots\mu_k}\mathrm{d}P_{\mu_1\cdots\mu_k},
\end{equation}
of which \((P_i, B^{\mu_1\cdots\mu_k})\) is the local coordinate of \(\Theta_J\in T^{*}J^{\infty}_y(E,\mathbb{P})\). Since that \cite{saunders_1989} 
\begin{equation}\label{eq-3-13}
\mathrm{d}P_{\mu_1\cdots\mu_k}=P_{\mu_1\cdots\mu_ki}\mathrm{d}y^i,
\quad 0\le k \le \infty,
\end{equation}
and subsequently, for \(k=0\), 
\begin{equation}\label{eq-3-14}
\mathrm{d}P=P_i\mathrm{d}y^i=\dot{P}\mathrm{d}t.
\end{equation}
For equation \eqref{eq-3-14}, there exists an isomorphism \(\varphi\) between \(T^{*}J^{\infty}_y(E,\mathbb{P})\) and \(T^{*}_y(E\times \mathbb{R})\) given by
\begin{equation}\label{eq-3-15}
\begin{split}
\varphi: T^{*}J^{\infty}_y(E,\mathbb{P})\to T^{*}_y(E\times \mathbb{R}),\;\;
\left\{
\begin{aligned}
P_i\mathrm{d}y^i&\mapsto\dot{P}\mathrm{d}t,\\
\sum_{0\le k\le \infty}\sum_{\mu_1<\cdots<\mu_k} B^{\mu_1\cdots\mu_k}\mathrm{d}P_{\mu_1\cdots\mu_k}&\mapsto \wp_i\mathrm{d}y^i.\\
\end{aligned}
\right.
\end{split}
\end{equation}
Obviously, \(\varphi\) is one-to-one and onto mapping that preserves contact structures. By the action of isomorphism defined by \(\varphi\), the contact 1-form \(\Theta_J \in T^{*}J^{\infty}(E,\mathbb{P})\) can be mapped to \(\Theta \in T^*(E\times \mathbb{R})\) as follows:
\begin{equation}\label{eq-3-16}
\begin{split}
	\Theta=\varphi\circ \Theta_J
	&=\mathcal{H}\mathrm{d}t-\wp_i\mathrm{d}y^i,
\end{split}
\end{equation}
where
\begin{equation}\label{eq-3-17}
\mathcal{H}=\dot{P}=\frac{\mathrm{d}P(y(t))}{\mathrm{d}t}.
\end{equation}

The 1-form \(\Theta = \varphi(\Theta_J)\) defines a canonical contact form on the extended phase space \(\mathcal{E} = T^{*}(E \times \mathbb{R})\) \cite{ArnoldVI1989, FOUNDATIONOFMECH, anold2010}, thereby constituting a contact manifold \((\mathcal{E},\Theta)\) of dimension \(2n+1\), as illustrated in Figure \ref{Fig:Fig-2}. 
It shows that \(\pi: TE \to E\) defines a tangent bundle on the vector space \(E\); \(\Theta=\varphi\circ \Theta_J: E\to T^*E\times\mathbb{R}\) and \(\sigma \circ \Theta_J=Id_E \); \(\varphi\) is defined in equation \eqref{eq-3-15}; \(\psi=\sharp\circ \pi_1\), of which \(\pi_1:T^*E\times \mathbb{R}\to T^*E\) and \(\sharp: T^*E\to TE\); \(\varphi\circ \Theta_J\) can be considered as a section for cotangent bundle \(\tau: \mathcal{E}\to E\). Apparently, this contact 1-form \(\Theta\) is equivalent to the connection 1-form \(\Theta_{\wp}\) upon substituting \(\mathrm{d}P = \mathcal{H}\mathrm{d}t\). This equivalence underscores that the horizontal subspace \(H\) of \(TJ^{\infty}(E, \mathbb{P})\), as determined by the connection 1-form, is isomorphic to the kernel of the contact 1-form. This insight is crucial for comprehending the statistical characteristics of stochastic processes. 

The contact form induces a 2-form \(\mathrm{d}\Theta \in \bigwedge^2 T^{*}\mathcal{E}\), which gives rise to a non-vanishing volume form \(\Theta\wedge(\mathrm{d}\Theta)^n \neq 0\) (refer to Lemma \ref{lem-1}). The 2-form \(\mathrm{d}\Theta: T\mathcal{E}\times T\mathcal{E}\to \mathbb{R}\) is expressed as: 
\begin{equation}\label{eq-3-18}
	\mathrm{d}\Theta=\mathrm{d}\mathcal{H}\wedge\mathrm{d}t-\mathrm{d}\wp_i\wedge\mathrm{d}y^i.
\end{equation}
With \(\Theta\) and \(\mathrm{d}\Theta\), the tangent space \(T\mathcal{E}_{(t,y,\wp)}\) of \(\mathcal{E}_{(t,y,\wp)}\) can be decomposed as a Whitney sum of the kernels of \(\Theta\) and \(\mathrm{d}\Theta\):
\begin{equation}\label{eq-3-19}
	T\mathcal{E}_{(t,y,\wp)}=\text{ker}(\Theta_{(t,y,\wp)})\oplus \text{ker}(\mathrm{d}\Theta_{(t,y,\wp)}).
\end{equation}

The decomposition of \(T\mathcal{E}_{(t,y,\wp)}\) by equation \eqref{eq-3-19} is a fundamental property that characterizes contact structures. It indicates that the contact form \(\Theta\) induces a splitting of the tangent space \(T\mathcal{E}_{(t,y,\wp)}\) at each point of the manifold into two complementary subspaces: the contact distribution \(\text{ker}(\Theta)\) and its orthogonal complement concerning the form \(\mathrm{d}\Theta\) induced by \(\Theta\). The decomposition in equation \eqref{eq-3-19} is significant because it allows for defining a unique vector field that is everywhere tangent to \(\text{ker}(\mathrm{d}\Theta)\) and orthogonal to \(\text{ker}(\Theta)\) in the sense of contact geometry. The vector field is fundamental for understanding the geometric properties of stochastic vector bundles: the evolution of the probability distribution \(P\) is determined by the curve of \(\text{ker}(\Theta)\); conversely, the curves of \(\text{ker}(\mathrm{d}\Theta)\) outline the evolution of the contact manifold \((\mathcal{E}, \Theta)\).

\begin{figure}[h!]
\centering
\begin{tikzcd}[ column sep=12em, row sep=13em]
TE
	\arrow{d}{\pi}
&\mathcal{E}= T^*E\times\mathbb{R}
	\arrow[swap]{l}{\psi}
	\arrow[yshift=-0.3ex, sloped, swap]{ld}{\tau}
\\
E
	\arrow[yshift=0.3ex]{r}{\Theta_J}
	\arrow[yshift=0.3ex, sloped]{ru}{\varphi\circ \Theta_J}	
&T^{*}J^{\infty}(E,\mathbb{P})
	\arrow{u}{\varphi}
	\arrow[yshift=-0.3ex]{l}{\sigma}
\end{tikzcd}
\caption{The commutative diagram for contact structure defined on \(J^{\infty}(E,\mathbb{P})\). \(\pi: TE \to E\) defines a tangent bundle on the vector space \(E\); \(\Theta=\varphi\circ \Theta_J: E\to T^*E\times\mathbb{R}\) and \(\sigma \circ \Theta_J=Id_E \); \(\varphi\) is defined in equation \eqref{eq-3-15}; \(\psi=\sharp\circ \pi_1\), of which \(\pi_1:T^*E\times \mathbb{R}\to T^*E\) and \(\sharp: T^*E\to TE\); \(\varphi\circ \Theta_J\) can be considered as a section for cotangent bundle \(\tau: \mathcal{E}\to E\).}
\label{Fig:Fig-2}
\end{figure}

What is most significant is the non-integrability of \(\text{ker}(\Theta)\). It highlights the close relationship between the distinctive characteristics of contact geometry and stochastic vector bundles. The kinetic equation derived in \cite{Zhong_2024, ZHONG-2022-KINETICEQAUTION} is path-dependent, as it is based on the cumulants of the system's state transition paths \(S\) as given in equation \eqref{eq-3-3}, implying that different paths result in different state possibilities. This leads to the possibility that the system can end up in different final states, \(P(y_C)\) and \(P(y_{C^{\prime}})\), even if it starts from the same initial state \(P(y_A)\) but follows different paths with different intermediate states \(P(y_B)\) and \(P(y_{B^{\prime}})\), as illustrated in Figure \ref{Fig:Fig-3}. Geometrically, this is explained by noting that the curve represented by the kinetic equation \eqref{eq-3-8} lies on \(\text{ker}(\Theta)\), which is completely non-integrable.

\begin{figure}[h!]
\centering

\tikzset{every picture/.style={line width=0.75pt}} 

\begin{tikzpicture}[x=0.85pt,y=0.85pt,yscale=-1,xscale=1]

\draw    (194,108) .. controls (234,78) and (303,100) .. (343,70) ;
\draw [shift={(274.22,88.58)}, rotate = 175.76] [fill={rgb, 255:red, 0; green, 0; blue, 0 }  ][line width=0.08]  [draw opacity=0] (10.72,-5.15) -- (0,0) -- (10.72,5.15) -- (7.12,0) -- cycle    ;
\draw    (194,108) .. controls (234,78) and (293,202) .. (290,238) ;
\draw [shift={(263.77,157.99)}, rotate = 240.57] [fill={rgb, 255:red, 0; green, 0; blue, 0 }  ][line width=0.08]  [draw opacity=0] (10.72,-5.15) -- (0,0) -- (10.72,5.15) -- (7.12,0) -- cycle    ;
\draw    (290,238) .. controls (310,190) and (396,177) .. (415,167) ;
\draw [shift={(350.5,189.21)}, rotate = 157.06] [fill={rgb, 255:red, 0; green, 0; blue, 0 }  ][line width=0.08]  [draw opacity=0] (10.72,-5.15) -- (0,0) -- (10.72,5.15) -- (7.12,0) -- cycle    ;
\draw    (343,70) .. controls (367,113) and (400,128) .. (416,124) ;
\draw [shift={(376.89,110.65)}, rotate = 219.19] [fill={rgb, 255:red, 0; green, 0; blue, 0 }  ][line width=0.08]  [draw opacity=0] (10.72,-5.15) -- (0,0) -- (10.72,5.15) -- (7.12,0) -- cycle    ;
\draw  [dash pattern={on 0.84pt off 2.51pt}]  (343,70) .. controls (341,127) and (404,143) .. (415,167) ;
\draw  [dash pattern={on 4.5pt off 4.5pt}]  (416,124) -- (415,167) ;
\draw   (132,109) .. controls (132,104.58) and (135.58,101) .. (140,101) .. controls (144.42,101) and (148,104.58) .. (148,109) .. controls (148,113.42) and (144.42,117) .. (140,117) .. controls (135.58,117) and (132,113.42) .. (132,109) -- cycle ; \draw   (136.48,106.28) .. controls (136.48,105.84) and (136.84,105.48) .. (137.28,105.48) .. controls (137.72,105.48) and (138.08,105.84) .. (138.08,106.28) .. controls (138.08,106.72) and (137.72,107.08) .. (137.28,107.08) .. controls (136.84,107.08) and (136.48,106.72) .. (136.48,106.28) -- cycle ; \draw   (141.92,106.28) .. controls (141.92,105.84) and (142.28,105.48) .. (142.72,105.48) .. controls (143.16,105.48) and (143.52,105.84) .. (143.52,106.28) .. controls (143.52,106.72) and (143.16,107.08) .. (142.72,107.08) .. controls (142.28,107.08) and (141.92,106.72) .. (141.92,106.28) -- cycle ; \draw   (136,112.2) .. controls (138.67,114.33) and (141.33,114.33) .. (144,112.2) ;
\draw  [fill={rgb, 255:red, 126; green, 211; blue, 33 }  ,fill opacity=1 ] (447,202.08) .. controls (447,197.62) and (451.03,194) .. (456,194) .. controls (460.97,194) and (465,197.62) .. (465,202.08) .. controls (465,206.54) and (460.97,210.15) .. (456,210.15) .. controls (451.03,210.15) and (447,206.54) .. (447,202.08) -- cycle ; \draw  [fill={rgb, 255:red, 126; green, 211; blue, 33 }  ,fill opacity=1 ] (452.04,199.33) .. controls (452.04,198.88) and (452.44,198.52) .. (452.94,198.52) .. controls (453.44,198.52) and (453.84,198.88) .. (453.84,199.33) .. controls (453.84,199.78) and (453.44,200.14) .. (452.94,200.14) .. controls (452.44,200.14) and (452.04,199.78) .. (452.04,199.33) -- cycle ; \draw  [fill={rgb, 255:red, 126; green, 211; blue, 33 }  ,fill opacity=1 ] (458.16,199.33) .. controls (458.16,198.88) and (458.56,198.52) .. (459.06,198.52) .. controls (459.56,198.52) and (459.96,198.88) .. (459.96,199.33) .. controls (459.96,199.78) and (459.56,200.14) .. (459.06,200.14) .. controls (458.56,200.14) and (458.16,199.78) .. (458.16,199.33) -- cycle ; \draw   (451.5,205.31) .. controls (454.5,207.46) and (457.5,207.46) .. (460.5,205.31) ;
\draw  [fill={rgb, 255:red, 208; green, 2; blue, 27 }  ,fill opacity=0.62 ] (330,24) .. controls (330,19.58) and (333.58,16) .. (338,16) .. controls (342.42,16) and (346,19.58) .. (346,24) .. controls (346,28.42) and (342.42,32) .. (338,32) .. controls (333.58,32) and (330,28.42) .. (330,24) -- cycle ; \draw  [fill={rgb, 255:red, 208; green, 2; blue, 27 }  ,fill opacity=0.62 ] (334.48,21.28) .. controls (334.48,20.84) and (334.84,20.48) .. (335.28,20.48) .. controls (335.72,20.48) and (336.08,20.84) .. (336.08,21.28) .. controls (336.08,21.72) and (335.72,22.08) .. (335.28,22.08) .. controls (334.84,22.08) and (334.48,21.72) .. (334.48,21.28) -- cycle ; \draw  [fill={rgb, 255:red, 208; green, 2; blue, 27 }  ,fill opacity=0.62 ] (339.92,21.28) .. controls (339.92,20.84) and (340.28,20.48) .. (340.72,20.48) .. controls (341.16,20.48) and (341.52,20.84) .. (341.52,21.28) .. controls (341.52,21.72) and (341.16,22.08) .. (340.72,22.08) .. controls (340.28,22.08) and (339.92,21.72) .. (339.92,21.28) -- cycle ; \draw   (334,27.2) .. controls (336.67,29.33) and (339.33,29.33) .. (342,27.2) ;
\draw  [fill={rgb, 255:red, 208; green, 2; blue, 27 }  ,fill opacity=1 ] (448,88) .. controls (448,83.58) and (451.58,80) .. (456,80) .. controls (460.42,80) and (464,83.58) .. (464,88) .. controls (464,92.42) and (460.42,96) .. (456,96) .. controls (451.58,96) and (448,92.42) .. (448,88) -- cycle ; \draw  [fill={rgb, 255:red, 208; green, 2; blue, 27 }  ,fill opacity=1 ] (452.48,85.28) .. controls (452.48,84.84) and (452.84,84.48) .. (453.28,84.48) .. controls (453.72,84.48) and (454.08,84.84) .. (454.08,85.28) .. controls (454.08,85.72) and (453.72,86.08) .. (453.28,86.08) .. controls (452.84,86.08) and (452.48,85.72) .. (452.48,85.28) -- cycle ; \draw  [fill={rgb, 255:red, 208; green, 2; blue, 27 }  ,fill opacity=1 ] (457.92,85.28) .. controls (457.92,84.84) and (458.28,84.48) .. (458.72,84.48) .. controls (459.16,84.48) and (459.52,84.84) .. (459.52,85.28) .. controls (459.52,85.72) and (459.16,86.08) .. (458.72,86.08) .. controls (458.28,86.08) and (457.92,85.72) .. (457.92,85.28) -- cycle ; \draw   (452,91.2) .. controls (454.67,93.33) and (457.33,93.33) .. (460,91.2) ;
\draw  [fill={rgb, 255:red, 126; green, 211; blue, 33 }  ,fill opacity=0.5 ] (287,274.08) .. controls (287,269.62) and (291.03,266) .. (296,266) .. controls (300.97,266) and (305,269.62) .. (305,274.08) .. controls (305,278.54) and (300.97,282.15) .. (296,282.15) .. controls (291.03,282.15) and (287,278.54) .. (287,274.08) -- cycle ; \draw  [fill={rgb, 255:red, 126; green, 211; blue, 33 }  ,fill opacity=0.5 ] (292.04,271.33) .. controls (292.04,270.88) and (292.44,270.52) .. (292.94,270.52) .. controls (293.44,270.52) and (293.84,270.88) .. (293.84,271.33) .. controls (293.84,271.78) and (293.44,272.14) .. (292.94,272.14) .. controls (292.44,272.14) and (292.04,271.78) .. (292.04,271.33) -- cycle ; \draw  [fill={rgb, 255:red, 126; green, 211; blue, 33 }  ,fill opacity=0.5 ] (298.16,271.33) .. controls (298.16,270.88) and (298.56,270.52) .. (299.06,270.52) .. controls (299.56,270.52) and (299.96,270.88) .. (299.96,271.33) .. controls (299.96,271.78) and (299.56,272.14) .. (299.06,272.14) .. controls (298.56,272.14) and (298.16,271.78) .. (298.16,271.33) -- cycle ; \draw   (291.5,277.31) .. controls (294.5,279.46) and (297.5,279.46) .. (300.5,277.31) ;
\draw    (283.62,117.01) -- (319.5,106.07) ;
\draw [shift={(321.41,105.48)}, rotate = 163.04] [color={rgb, 255:red, 0; green, 0; blue, 0 }  ][line width=0.75]    (10.93,-3.29) .. controls (6.95,-1.4) and (3.31,-0.3) .. (0,0) .. controls (3.31,0.3) and (6.95,1.4) .. (10.93,3.29)   ;
\draw    (283.62,117.01) -- (298.09,145.22) ;
\draw [shift={(299,147)}, rotate = 242.85] [color={rgb, 255:red, 0; green, 0; blue, 0 }  ][line width=0.75]    (10.93,-3.29) .. controls (6.95,-1.4) and (3.31,-0.3) .. (0,0) .. controls (3.31,0.3) and (6.95,1.4) .. (10.93,3.29)   ;

\draw (150,100.4) node [anchor=north west][inner sep=0.75pt]    {$P( y_A)$};
\draw (279,243.4) node [anchor=north west][inner sep=0.75pt]    {$P( y_B)$};
\draw (417,170.4) node [anchor=north west][inner sep=0.75pt]    {$P( y_C)$};
\draw (418,120.6) node [anchor=south west] [inner sep=0.75pt]    {$P\left( y_{C'}\right)$};
\draw (319,39.4) node [anchor=north west][inner sep=0.75pt]    {$P\left( y_{B^{'}} \right)$};
\draw (283,177.4) node [anchor=north west][inner sep=0.75pt]    {$\mathrm{k} er( \Theta )$};
\draw (278.12,138.01) node [anchor=north west][inner sep=0.75pt]  [rotate=-342.22]  {$y$};
\draw (320.35,97.71) node [anchor=north west][inner sep=0.75pt]  [rotate=-342.22]  {$P$};

\end{tikzpicture}

\caption{Schemetic diagram of the non-integrable stochastic system. It indicates that even starting at the same state \(P(y_A)\) but with different paths with different intermediate states of \(P(y_B)\) and \(P(y_{B^{\prime})}\), the system will have different final probabilities \(P(y_C)\) and \(P(y_{C^{\prime}})\), respectively.}
\label{Fig:Fig-3}
\end{figure}

In conclusion, the 1-form \(\Theta\), as detailed in equation \eqref{eq-3-16}, along with its exterior derivative \(\mathrm{d}\Theta\) given in equation \eqref{eq-3-18}, is a typical contact form. It satisfies the properties of non-degeneracy, the presence of a non-vanishing volume form, and maximal non-integrability. Therefore, it is evident that stochastic vector bundles are equipped with well-defined contact structures \((\mathcal{E}, \Theta)\). This result not only confirms the theoretical foundations of these structures but also paves the way for the subsequent application of contact geometry in analyzing the evolution of stochastic vector bundles.

\section{Contact Dynamics }\label{sec-4}
This section discusses the contact dynamics of stochastic vector bundles through the procedure for deriving the contact dynamical equations. We begin with the following lemma. 
\vspace{1em}
\begin{lemma}[Contact Dynamical Equations of Stochastic Vector Bundles]\label{lem-2}
On the stochastic contact manifold \((\mathcal{E}, \Theta)\), the contact 1-form \(\Theta\) and the smooth constraint function \(\varepsilon: \mathcal{E} \rightarrow \mathbb{R}\) uniquely determine a vector field \(\mathfrak{X}(\mathcal{E}) = \{X_{\mathcal{H}} \mid X_{\mathcal{H}} \in \ker (\mathrm{d}\Theta) \subset T\mathcal{E}\}\). This vector field satisfies:
\begin{enumerate}[label=(\roman*)]
    \item \(\iota_{X_{\mathcal{H}}}\Theta = -\varepsilon\);
    \item \(\iota_{X_{\mathcal{H}}} (\mathrm{d}\Theta) = 0\);
    \item \(L_{X_{\mathcal{H}}}\Theta = 0\).
\end{enumerate}
The vector field \(\mathfrak{X}(\mathcal{E})\) generates a flow that preserves the contact structure and leads to the following equations:
\begin{equation*}
\left\{
\begin{aligned}
	\frac{\mathrm{d}y^i}{\mathrm{d}t} &=\:\:\; \{y^i, \mathcal{H}\}, \\
	\frac{\mathrm{d}\wp_i}{\mathrm{d}t} &=\:\:\; \{\wp_i, \mathcal{H}\}, \\
	\frac{\partial \varepsilon}{\partial t} &= -\{\varepsilon, \;\  \mathcal{H}\}.
\end{aligned}
\right.
\end{equation*}

\end{lemma}
\vspace{1em}
\begin{proof}

Let \(X_{\mathcal{H}}\) represent the vector tangent to \(\mathcal{E}\), i.e., \(X_{\mathcal{H}} \in T\mathcal{E}\). Furthermore, we restrict it to constitute a special vector field \(\mathfrak{X}(\mathcal{E}) = \{X_{\mathcal{H}} \mid X_{\mathcal{H}} \in \mathrm{ker} (\mathrm{d}\Theta) \subset T\mathcal{E}\}\) and thus it generates a flow that encapsulates the dynamic behavior of the contact manifold. This flow is unique because it transports the contact structure in \(\mathcal{E}\) without losing its nature. Preserving the contact structure ensures the system's evolution adheres to certain geometric and dynamical constraints \cite{ArnoldVI1989}. Understanding how the flow preserves the structure of the contact manifold \((\mathcal{E}, \Theta)\) is necessary for obtaining the qualitative characteristics of the system's motion, as it governs the evolution of the system's trajectories over time while maintaining the contact structure, which is a constraint imposed by the underlying geometry.

The contact structure defined by the 1-form \(\Theta\) is classified as the infinitesimal contact system if, and only if, it exhibits invariance along \(X_{\mathcal{H}}\). It requires that \(\Theta\) is  invariant along the path determined by \(X_{\mathcal{H}} \), i.e., 
\begin{equation}\label{eq-4-1}
	L_{X_{\mathcal{H}}}\Theta=\mathrm{d}(\iota_{X_{\mathcal{H}}}\Theta)+\iota_{X_{\mathcal{H}}}\mathrm{d}\Theta=0.
\end{equation}
Here, we assume that \(X_{\mathcal{H}} \in \mathrm{ker}(\mathrm{d}\Theta)\subset T\mathcal{E}\) has the form in terms of the local coordinate \((t, y^i, \wp_i)\in \mathcal{E}\) as 
\begin{equation}\label{eq-4-2}
 X_{\mathcal{H}}=A^t\frac{\partial}{\partial t}+B^i\frac{\partial }{\partial y^i}+C_i\frac{\partial }{\partial \wp_i},
\end{equation}
where \(A^t\), \(B^i\), and \(C_i\) are parameters to be determined.

Equation \eqref{eq-4-1} shows that the vector \(X_{\mathcal{H}}\) must satisfy, firstly, 
\begin{equation}\label{eq-4-3}
\iota_{X_{\mathcal{H}}}\mathrm{d}\Theta
	=0.	
\end{equation}
This condition results in  
\begin{equation}\label{eq-4-4}
\begin{split}
	0&=\iota_{X_{\mathcal{H}}}\mathrm{d}\Theta\\
	&=-\iota_{X_{\mathcal{H}}}(\mathrm{d}\wp_i\wedge\mathrm{d}y^i-\mathrm{d}\mathcal{H}\wedge\mathrm{d}t)\\
	&=-X_{\mathcal{H}}(\wp_i)\mathrm{d}y^i+X_{\mathcal{H}}(y^i)\mathrm{d}\wp_i+X_{\mathcal{H}}(\mathcal{H})\mathrm{d}t-X_{\mathcal{H}}(t)\mathrm{d}\mathcal{H}
	\\
	&=-C_i\mathrm{d}y^i+B^i\mathrm{d}\wp_i
	+X_{\mathcal{H}}(\mathcal{H})\mathrm{d}t
	-A^t\mathrm{d}\mathcal{H}.	
\end{split}
\end{equation}
Since 
\begin{equation}\label{eq-4-5}
\iota_{X_{\mathcal{H}}}\mathrm{d}t=A^t=1,
\end{equation}
which ensures that the system evolves within a unified temporal framework, it is deduced from equation \eqref{eq-4-4} that  
\begin{equation}\label{eq-4-6}
\frac{\mathrm{d}\mathcal{H}}{\mathrm{d}t}=\frac{\partial \mathcal{H}}{\partial t},
\end{equation}
which implies that \(\mathcal{H}\) itself explicitly depends on time, and furthermore, 
\begin{equation}\label{eq-4-7}
	 -C^i\mathrm{d}\wp_i+ D_i\mathrm{d}y^i
	 =-\frac{\partial \mathcal{H}}{\partial \wp_i}\mathrm{d}\wp_i-\frac{\partial \mathcal{H}}{\partial y^i}\mathrm{d}y_i.
\end{equation}
Comparing the terms with \(\mathrm{d}\wp_i\) and \(\mathrm{d}y^i\) on both sides of equation \eqref{eq-4-7} results in 
\begin{subequations}\label{eq-4-8}
	\begin{align}[left = \empheqlbrace\,]
	C^i&=\:\:\;\frac{\partial \mathcal{H}}{\partial \wp_i}\label{eq-4-8a},\\
	D^i&=-\frac{\partial \mathcal{H}}{\partial y^i}.\label{eq-4-8b}	
	\end{align}
\end{subequations}
Consequently, the vector \(X_{\mathcal{H}}\) is given by
\begin{equation}\label{eq-4-9}
	X_{\mathcal{H}}=\frac{\partial}{\partial t}
	+\frac{\partial \mathcal{H}}{\partial \wp_i}\frac{\partial }{\partial y^i}-\frac{\partial \mathcal{H}}{\partial y^i}\frac{\partial }{\partial \wp_i}.
\end{equation}
Moreover, equation \eqref{eq-4-9} leads to a set of contact dynamical equations  as follows:
\begin{subequations}\label{eq-4-10}
	\begin{align}[left = \empheqlbrace\,]
		\frac{\mathrm{d}y^i}{\mathrm{d}t}
		&=\:\:\;\frac{\partial \mathcal{H}}{\partial \wp_i},\label{eq-4-10a}\\
		\frac{\mathrm{d}\wp_i}{\mathrm{d}t}
		&=-\frac{\partial \mathcal{H}}{\partial y^i},\label{eq-4-10b}\\
		\frac{\mathrm{d}\mathcal{H}}{\mathrm{d}t}
		&=\:\:\;\frac{\partial\mathcal{H}}{\partial t}.\label{eq-4-10c}
	\end{align}
\end{subequations}
which depicts how the system \((y^i, \wp_i, \mathcal{H})\) evolves over time.

Equation \eqref{eq-4-10c} only shows that \(\mathcal{H}\) is time-dependent, while determining how it changes over time requires another condition. This condition is also given by equation \eqref{eq-4-1}, which is 
\begin{equation}\label{eq-4-11}
\mathrm{d}(\iota_{X_{\mathcal{H}}}\Theta)=0.	
\end{equation}
Because \(\Theta\) is a 1-form, $\iota_{X_{\mathcal{H}}}\Theta$ must be a 0-form, i.e., a function, or a constant. Also, because \(X_{\mathcal{H}}\notin \text{ker}(\Theta)\), it is required that $\iota_{X_{\mathcal{H}}}\Theta\neq 0$. For example, when we let that 
\begin{equation}\label{eq-4-12}
	\iota_{X_{\mathcal{H}}}\Theta=1,
\end{equation}
it leads to Reeb's contact vector field. Reeb's vector is special because it has a unit length in the direction of \(\Theta\). Certainly, there are other choices \cite{Bravetti-CONTACTHAMILTONIAN}. For a general case of stochastic jet bundles, we assumed that there is a smooth function \(\varepsilon \in C^{\infty}\) that \(\varepsilon: \mathcal{E}\to \mathbb{R}\) and \(\varepsilon \neq 0\),  termed as the  constraint function, satisfying that  
\begin{equation}\label{eq-4-13}
	\iota_{X_{\mathcal{H}}}\Theta=-\varepsilon(t,y(t),\wp(t)),
\end{equation}
and
\begin{equation}\label{eq-4-14}
\begin{split}
	\mathrm{d}\varepsilon
	=0.
\end{split}
\end{equation}

Equation \eqref{eq-4-13} results in 
\begin{equation}\label{eq-4-15}
	\varepsilon=\wp_i\dot{y}^i-\mathcal{H},
\end{equation}
whereas equation \eqref{eq-4-14} is equivalent to
\begin{equation}\label{eq-4-16}
\begin{split}
	\frac{\mathrm{d}\varepsilon}{\mathrm{d}t}
	&=
	\frac{\partial \varepsilon}{\partial t}
	+\frac{\mathrm{d}y^i}{\mathrm{d}t}\frac{\partial \varepsilon}{\partial y^i}
	+\frac{\mathrm{d}\wp_i}{\mathrm{d}t}\frac{\partial \varepsilon}{\partial \wp_i}\\
	&=X_{\mathcal{H}}\varepsilon
	=L_{X_{\mathcal{H}}}\varepsilon=0,
\end{split}
\end{equation}
with $L_{X_{\mathcal{H}}}$ denoting the Lie derivative induced by the vector \(X_{\mathcal{H}}\).  By equation \eqref{eq-4-15}, we can determine \(\tfrac{\partial \mathcal{H}}{\partial t}\) in equation \eqref{eq-4-10c} as below:
\begin{equation}\label{eq-4-17}
	L_{X_{\mathcal{H}}}\varepsilon=
	-\left[\frac{\partial \mathcal{H}}{\partial t}-L_{X_{\mathcal{H}}}\left(\wp_i\dot{y}^i\right)\right]=0.
\end{equation}

Because the contact dynamical equations resemble Hamilton's equations, we can, correspondingly, adopt the Poisson brackets  
\begin{equation}\label{eq-4-24}
	\{F, \mathcal{H}\}
	\equiv 
	\frac{\partial F}{\partial y^i}\frac{\partial \mathcal{H}}{\partial \wp_i}
	-\frac{\partial F}{\partial \wp_i}\frac{\partial \mathcal{H}}{\partial y^i}, 
\end{equation} 
and thus any conservative variable \(F\) transported in this system will observe the law 
\begin{equation}\label{eq-4-25}
	\frac{\mathrm{d}F}{\mathrm{d}t}=\frac{\partial F}{\partial t}+\{F, \mathcal{H}\}=0.
\end{equation} 
Using the definition of Poisson brackets as in equation \eqref{eq-4-24}, it can be verified that 
\begin{subequations}\label{eq-4-26}
	\begin{align}[left = \empheqlbrace\,]
		\{y^i, y^j\}   &=0, \label{eq-4-26a}\\
		\{\wp_i,\wp_j\}&=0, \label{eq-4-26b}\\
		\{y^i,\wp_j\}  &=\tensor{\delta}{^i_j}.\label{eq-4-26c}
	\end{align}	
\end{subequations}
Subsequently, the contact dynamical equations, when expressed in terms of Poisson brackets, are
\begin{subequations}\label{eq-4-27}
	\begin{align}[left = \empheqlbrace\,]
		\frac{\mathrm{d}y^i}{\mathrm{d}t}&
		=\{y^i, \mathcal{H}\},\label{eq-4-27a}\\
		\frac{\mathrm{d}\wp_i}{\mathrm{d}t}&
		=\{\wp_i, \mathcal{H}\},\label{eq-4-27b}\\
		\frac{\mathrm{d}\mathcal{H}}{\mathrm{d}t}&
		=\frac{\partial \mathcal{H}}{\partial t}=L_{X_{\mathcal{H}}}\left(\wp_i\dot{y}^i\right)
		=\frac{\partial \wp_i\dot{y}^i}{\partial t}+\{\wp_i\dot{y}^i,\mathcal{H}\}.
		\label{eq-4-27c}
	\end{align}
\end{subequations}
Furthermore, because \(\varepsilon=\wp_i\dot{y}^i-\mathcal{H}\), we had that 
\begin{equation}\label{eq-4-28}
\frac{\partial \varepsilon}{\partial t}
=-\{\wp_i\dot{y}^i,\mathcal{H}\}
=-\{\varepsilon+\mathcal{H},\mathcal{H}\}
=-\{\varepsilon,\mathcal{H}\},
\end{equation}
which is equivalent to equations \eqref{eq-4-16} and \eqref{eq-4-27c}. For the convenience of applications, \eqref{eq-4-27} can be written as below:
\begin{subequations}\label{eq-4-29}
	\begin{align}[left = \empheqlbrace\,]
		\frac{\mathrm{d}y^i}{\mathrm{d}t}&
		=\:\:\;\{y^i, \mathcal{H}\},\label{eq-4-29a}\\
		\frac{\mathrm{d}\wp_i}{\mathrm{d}t}&
		=\:\:\;\{\wp_i, \mathcal{H}\},\label{eq-4-29b}\\ 
		\frac{\partial \varepsilon}{\partial t}&
		=- \{ \varepsilon, \;\ \mathcal{H} \}.
		\label{eq-4-29c}
	\end{align}
\end{subequations}

The above analysis proves the existence of such a vector field \(X_{\mathcal{H}}\) along the paths represented by equations \eqref{eq-4-29}; the proof of its uniqueness is classical; for example,   
Theorem 2.1.2 of \cite{FOUNDATIONOFMECH}. 
\\
\end{proof}

The contact dynamical equations for stochastic vector bundles, encapsulated in equations \eqref{eq-4-29}, can be used to calculate the temporal evolution of a system characterized by the triplet of variables \((y, \wp, \varepsilon)\). These equations, defined in terms of Poisson brackets with a Hamiltonian-like function \( \mathcal{H} \), reveal how each component of the system changes over time. The variables \( y^i \) and \( \wp_i \) follow distinct evolution rules governed by their respective Poisson brackets with \( \mathcal{H} \), indicating a structured interplay between the state of the system and its state distribution \(P(y)\). Meanwhile, the time derivative of the constraint \( \varepsilon \) is inversely related to its Poisson bracket with \( \mathcal{H} \), suggesting a balance or conservation principle within the system. Central to the contact dynamical system is \( \mathcal{H} \), which measures the time rate of change of \( y \) in the sense of probability by a 1-form of the probability \( P \). When it is prescribed, the evolution of realizations \( y \in E \) of the random variable \( Y \), as well as \(\wp\) and \(\varepsilon\), is obtained.  

Equations \eqref{eq-4-29a} and \eqref{eq-4-29b} resemble Hamilton's canonical equations. However, essential differences were observed. Table \ref{Tab:tab-4-1} compares Hamilton's equations for deterministic systems, one of the outcomes of symplectic geometry, and the contact dynamical equations for stochastic vector bundles. They differ in several respects. In Hamilton's equations, the generalized coordinates \( y^i \) represent the configuration of the system, whereas \( \wp_i=\tfrac{\partial \mathcal{L}}{\partial \dot{y}^i} \) denotes the conjugate momenta related to the Lagrangian, defined as \(\mathcal{L}=\wp_i\dot{y}^i-\mathcal{H}\). The Hamiltonian \( \mathcal{H} =T+U\) represents the total energy of the system as the summation of the kinetic energy \(T\) and potential energy \(U\), whereas the Lagrangian \( \mathcal{L} \) is expressed as the difference between the kinetic and potential energies.

By contrast, this study redefines the variables for stochastic systems. Here, \( y^i \) is interpreted as a vector in phase space, representing a realization of probabilistic states. The conjugate momentum \( \wp_i \) is redefined as the probability flux, reflecting the flow of probability in phase space \( \mathcal{E} \). This is due to the fact that \(\wp_i\), as shown in equation \eqref{eq-A-18}, is a function of \(P_{\mu_1\cdots\mu_k}=\partial_{\mu_1\cdots\mu_k}P, \; (1\le k\le \infty)\), which represents flux of probability. The Hamiltonian is replaced by \( \mathrm{d}P(\dot{y}) \). The most important point is that this paper shifts the focus from deterministic to probabilistic, potentially non-conservative systems by redefining the fundamental variables and functions of Hamiltonian mechanics in terms of probabilistic quantities. This research differs from classical mechanics in that it encompasses more comprehensive phenomena, including stochastic and non-conservative ones. This new approach applies to various practical issues in which conventional deterministic Hamiltonian mechanics is inadequate.
\\

An illustrative example is presented below for a deeper insight into stochastic contact dynamics.

\vspace{1em}
\begin{example}[Stochastic version of Reeb's vector field] 
Solving the contact dynamics equations in phase space \(\mathcal{E}\) presents significant challenges due to the complexity of calculating the Poisson bracket, which involves gradients in phase space. However, in some special cases, analytical solutions exist. For example, one arises when \(\varepsilon=1\), leading to the stochastic version of Reeb's vector field. For the special case in \eqref{eq-4-29} where \(\varepsilon=1\), the system becomes
\begin{equation*}
\left\{
	\begin{aligned}
		\frac{\mathrm{d}y^i}{\mathrm{d}t} 
		&= \dot{y}^i,\\
		\frac{\mathrm{d}\wp_i}{\mathrm{d}t} 
		&= -\wp_j\frac{\partial \dot{y}^j}{\partial y^i}
		=-\wp_i\tensor{\delta}{^i_j}\frac{\partial \dot{y}^j}{\partial y^i}=-\wp_i \frac{\partial \dot{y}^j}{\partial y^j},\\ 
		\varepsilon &= 1.
	\end{aligned}
\right.
\end{equation*}
The above contact dynamical equations have an analytical solution as 
\begin{equation*}
\left\{
\begin{aligned}
y^i(t)&=\int\dot{y}^i(s)\mathrm{d}s,\\
\wp_i(t)&=\wp_i(0)\mathrm{e}^{-\int_0^t \frac{\partial \dot{y}^j}{\partial y^j} \mathrm{d}s},\\
\varepsilon&=1,
\end{aligned}
\right.
\end{equation*}
which represents a stochastic version of Reeb's field.

If we further assume that the system is divergence-free, indicating the absence of sinks or sources within the contact vector field, or more specifically, that the system states neither disappear nor appear, then it follows that \(\tfrac{\partial \dot{y}^j}{\partial y^j} = 0\), and thus 
\[
\wp_i(t)=\wp(0),
\]
implying that  \(\wp\) is a constant. Thus, the system further reduces to:
\begin{equation*}
\left\{
	\begin{aligned}
		y^i(t)&=\int\dot{y}^i(s)\mathrm{d}s, \\
		\wp_i&= \sum_{1\le k\le \infty}\sum_{\mu_1<\cdots<\mu_k}{B}^{\mu_1\cdots\mu_k}_iP_{\mu_1\cdots\mu_k}= C, \\
		\varepsilon&=1,
	\end{aligned}
\right.
\end{equation*}
where \(C\) is constant. This represents stable stochastic Reeb's field, specifically the probability flux of \(P\) observed in steady states. More precisely, if we consider that \(\wp\) as the section of \(\tau: \mathcal{E} \to E\), it yields:
\begin{equation*}
\begin{split}
\wp_i(y)
&=\sum_{1\le k\le \infty}\sum_{\mu_1<\cdots<\mu_k}{B}^{\mu_1\cdots\mu_k}_iP_{\mu_1\cdots\mu_k}(y)\\
&=\sum_{1\le k\le \infty}\sum_{\mu_1<\cdots<\mu_k}{B}^{\mu_1\cdots\mu_k}_i
\left(\frac{\partial }{\partial y^{\mu_1}}\otimes\cdots\otimes \frac{\partial }{\partial y^{\mu_k}}\right)P(y)
=C,
\end{split}
\end{equation*}
which is an infinite-order partial differential equation about \(P\). Usually, we set \(C=0\), its solution precisely gives the stable \(P(y)\), i.e., the section of \(\pi_E: \mathbb{P}\to E\). If it is truncated in the second order, we have that 
\[
\left(
-D^{\mu}\frac{\partial}{\partial y^{\mu}}
+
\frac{1}{2!}D^{\mu\nu}\frac{\partial}{\partial y^{\mu}}\otimes\frac{\partial}{\partial y^{\nu}}
\right)P(y)=0,
\]
gives the second-order kinetic equation for stable systems, 
with 
\[
B_i^{\mu}\mathrm{d}y^i(\dot{y})=-D^{\mu}
\] 
and 
\[
B_i^{\mu\nu}\mathrm{d}y^i(\dot{y})=\frac{1}{2!}D^{\mu\nu},
\]
given by \eqref{eq-A-16} in Appendix \ref{app-1}. 
\end{example}
\vspace{1em}

This example, while straightforward, is significant as it offers clear insight into the behavior of stochastic systems under constant constraints. Specifically, when the constant constraint \(\varepsilon=1\) is applied to the system, the contact dynamical equations that govern the interactions within the system reveal the fundamental properties of the stochastic vector bundles, particularly their stable probabilities. These stable probabilities are crucial because they reflect the long-term behavior of the system, indicating the likelihood of different states under specified constraints. The example underscores the importance of understanding the relationship between geometry and contact dynamics in analyzing the stochastic properties of vector bundles.
%
%
\begin{table}[htpb!]
\caption{Comparisons of Hamilton formalism for classical mechanics and contact dynamical equations of stochastic vector bundles}
\centering
\begin{tabularx}{\textwidth}{lXX}
\toprule
Variables	
& Hamilton Formalism    				
& Contact Dyanmics \\
\midrule  
$y^i$		
& Generalized coordinates of the system  \footnotemark[1]		
& Vector in phase space\\
$\wp_i$     
& Conjugate momentum   \footnotemark[2]     	
& Probability flux \footnotemark[3]\\
$\mathcal{H}$   
& Hamiltonian, total energy 	
&  $\mathrm{d}P(\dot{y})$, work done in changing  system states \\
$\mathcal{L}$	
& Lagrangian, $\mathcal{L}=\wp_i\dot{y}^i-\mathcal{H}$			
& -\\
\(\varepsilon\)
& -
&\(\varepsilon=\wp_i\dot{y}^i-\mathcal{H}\), Constraint as energy residual or net energy change in state transitions\\
\bottomrule
\end{tabularx}
\label{Tab:tab-4-1}
\footnotetext[1]{\(y^i\) is usually denoted as \(q^i\) in classical mechanics.}
\footnotetext[2]{\(\wp_i\) is usually denoted as \(p_i\) in classical mechanics.}
\footnotetext[3]{\(\wp_i\), as shown in equation \eqref{eq-A-18}, is a function of \(P_{\mu_1\cdots\mu_k}(y)=\partial_{\mu_1\cdots\mu_k}P, \; 1\le k\le \infty\), which represents flux of probability.}
\end{table}

\section{Least Constraint}\label{sec-5}
This section of the paper presents a theorem regarding the least constraint on the evolution of stochastic vector bundles. The theorem can be regarded as a synthesis of the research detailed in Sections \ref{sec-2}, \ref{sec-3}, and \ref{sec-4}, which act as foundational lemmas leading up to the theorem. 

The variable \(\varepsilon\) in equation \eqref{eq-4-15} appears to share a similar form to the Lagrangian \(\mathcal{L}\) of classical mechanics, which is the core of classical mechanics or symplectic geometry and is related to the least action principle. Consequently, a reasonable question arises: Does a similar principle exist for contact manifolds of stochastic vector bundles? Prior to addressing this question, it is essential to examine the nature of \(\varepsilon\) to distinguish it from \(\mathcal{L}\).

It can be found that critical differences exist. First, the Lagrangian must satisfy \(L=L(y,\dot{y})\) and \(\wp_i=\tfrac{\partial L}{\partial \dot{y}^i}\), whereas these conditions are unnecessary for \(\varepsilon\). Second, if we consider that \(0 \le P \le 1\) represents a normalized potential, then  
\begin{equation}\label{eq-4-18}
\mathcal{H} = \mathrm{d}P(\dot{y}) =\dot{y}^i  \frac{\mathrm{d}P}{\mathrm{d}y^i} 
\end{equation}
can be interpreted as the energy change due to a state transition from \(P(y)\) to \(P(y + \mathrm{d} y)\) along the path \(y(t)=y(0)+\int \dot{y}^i \mathrm{d}t\) within an infinitesimal time interval \(\mathrm{d}t\to 0\), rather than the total energy in classical mechanics. 

By analyzing the nature of \(\mathrm{d}P\) and \(\wp_i\dot{y}^i\), we reach the following observations:

(1) If \(\frac{\mathrm{d}P}{\mathrm{d}y}>0\), the system undergoes a transition from a state of lower potential energy to a higher one along \(y(t)=y(0)+\int \dot{y}^i \mathrm{d}t\). This indicates that an external influence or internal process provides power to the system, enabling it to overcome potential barriers or increase its energy content. Such a transition may arise from various mechanisms, including applying force, absorbing heat, or releasing stored energy within the system. The system dynamics in this scenario are characterized by a more disordered or less stable state, depending on the specific context of the system under examination.

(2) Conversely, when \( \frac{\mathrm{d}P}{\mathrm{d}y} < 0 \), the system transitions from a higher to a lower potential energy state along the path \(y(t)=y(0)+\int \dot{y}^i \mathrm{d}t\). This phenomenon typically occurs when a system expels energy to its environment either by performing work or by emitting heat. Internal dissipation also results in \( \frac{\mathrm{d}P}{\mathrm{d}y} < 0 \). Moving toward a lower potential energy state often correlates with a more ordered or stable condition, signifying a relaxation process or the natural tendency of the system to minimize energy and reach equilibrium.

(3) Additionally, \(\wp_i \dot{y}^i\) quantifies the work done by the probability flux \(\wp_i\) as it moves along the trajectory defined by \(y(t) = y(0) + \int \dot{y}^i \mathrm{d}t\). This study is directly related to the change in the system state and measures the energy transport that occurs during the transition. The probability flux \(\wp\) encapsulates the probability flow within the phase space of the system and is influenced by external forces and internal fluctuations, see \cite{Zhong_2024, ZHONG-2022-KINETICEQAUTION} or the explanations in Section \ref{app-1}. External forces, such as applied fields, gradients, or mechanical stresses, can alter the probability distribution and influence the system behavior. Internal fluctuations arising from the inherent agitation of the system constituents also play a significant role in determining the probability flux. These fluctuations manifest the microscopic dynamics of the system and can lead to random variations in its state.

Therefore, the interplay between the potential energy landscape, indicated by \(\tfrac{\mathrm{d}P}{\mathrm{d}y}\), and probability flux \(\wp_i \dot{y}^i\) provides a detailed picture of the system's energetics and dynamics. As a result, we have the following lemma: 
\vspace{1em}
\begin{lemma}[Least Constraint of Stochastic Vector Bundles]\label{lem-3}
The flow generated by the vector field $\mathfrak{X}(\mathcal{E})$ corresponds to the extremal paths for the action of the constraint function $\epsilon$, defined by
\[
\mathcal{S} = \int \epsilon \mathrm{d}t = -\int \Theta
\]
to reach its extreme. 
\end{lemma}
\vspace{1em}

\vspace{1em}
\begin{proof}

This lemma implies that the system's evolution is such that it minimizes the constraint \(\varepsilon\), which represents the residual energy or efficiency in converting work done by the probability flux into useful energy for state transitions, and simultaneously maximizes the variation in probabilities. The constraint \(\varepsilon\) can be understood from both the energy and constraint viewpoints:  

\textbf{(1) Energy aspect}. The variable \(\varepsilon = \wp_i \dot{y}^i - \mathrm{d}P(\dot{y})\) quantifies how the energy landscape of the system is altered as it evolves over time and measures the residual energy available to sustain further flows or transitions within the system. In other words, it indicates the extent to which the net energy input into the system (through the work done by the probability flux) exceeds the energy required to change the potential energy of the system. This residual or net energy change is crucial for understanding the ability of a system to maintain its dynamics such as continuous flows, oscillations, or other forms of motion. 

Moreover, \(\varepsilon\) can be seen as a measure of the system's efficiency in converting the work done by the probability flux into useful energy for state transition. A high \(|\varepsilon|\) value implies inefficiencies in energy conversion, whereas a low \(|\varepsilon|\) value suggests higher state-transition efficiencies. For instance, when energy is transferred from the flux kinetic energy to the potential energy, it is assumed that \(\varepsilon=\wp_i\dot{y}^i-\mathrm{d}P(\dot{y})>0\); a lower \(\varepsilon\) implies that energy is rapidly moved from the flux kinetic energy to the potential energy to increase the potential energy. Conversely, energy is transferred from the potential to the flux kinetic energy. 

Therefore,
\begin{equation}\label{eq-4-19}
\int \varepsilon\mathrm{d}t
=\int \left[\wp_i+\left(-\frac{\mathrm{d}P}{\mathrm{d}y^i}\right)\right]\mathrm{d}y^i(\dot{y})\mathrm{d}t
=\int\left[\wp_i+\left(-\frac{\mathrm{d}P}{\mathrm{d}y^i}\right)\right]\mathrm{d}y^i
\end{equation}
is reasonably interpreted as the residual energy or net energy change in state transitions, which measures the energy transfer efficiency. At the same time, \( \wp_i\dot{y}^i=\varepsilon+\mathrm{d}P(\dot{y}) \) can be regarded as the conservation of total energy, indicating that the net work done by the probability flux \( \wp \) is used to constrain the system represented by \(\varepsilon\) and change the system state represented by \( \mathrm{d}P(\dot{y}) \).

\textbf{(2) Constraint aspect}. Geometrically, \(\varepsilon\) is a constraint on the cantact system \((\mathcal{E}, \Theta)\). It can be found that Reeb's vector field is a special case of this study for \(\varepsilon=1\). Since that 
\begin{equation}\label{eq-4-21}
\frac{\mathrm{d}\varepsilon}{\mathrm{d}t}=L_{X_{\mathcal{H}}}\varepsilon=0,
\end{equation}
which indicates that the constraint is invariant along \(X_{\mathcal{H}}\). This property is significant because it maintains an invariant contact structure.

Furthermore, \(\varepsilon\) has another significant property. Let
\begin{equation}\label{eq-4-22}
	\mathcal{S}
	=\int \varepsilon \mathrm{d}t
	=\int (\wp_i\dot{y}^i-\mathcal{H}) \mathrm{d}t
	=\int \wp_i\mathrm{d}y^i-\mathcal{H}\mathrm{d}t
	=-\int \Theta, 
\end{equation}
the variation of $\mathcal{S}$ is given by
\begin{equation}\label{eq-4-23}
\begin{split}
	\delta \mathcal{S}=&
	\int \left[\mathrm{d}\wp_i+\left(\frac{\partial \mathcal{H}}{\partial y^i}\right)\mathrm{d}t  \right]\delta y^i
	+\int \left[-\mathrm{d}y^i+\left(\frac{\partial \mathcal{H}}{\partial \wp_i}\right)\mathrm{d}t  \right]\delta \wp_i\\
	&+\int \left[-\mathrm{d}\mathcal{H}+\left(\frac{\partial \mathcal{H}}{\partial t}\right)\mathrm{d}t  \right]\delta t+\int \mathrm{d}\left(\mathcal{H}\delta t-\wp_i\delta y^i\right).	
\end{split}
\end{equation}	
Since at the beginning and ending integration points \(\delta t=\delta y^i=0\), we found that \(\delta \mathcal{S}=0\) when substituting equation \eqref{eq-4-10} into it. Inversely, \(\delta \mathcal{S}=0\) leads to equation \eqref{eq-4-10}. Because equation \eqref{eq-4-13} is viewed geometrically as a constraint on the system, \(\delta \mathcal{S}=0\) illustrates that the contact dynamical equations represent the trajectories leading to minimum constraints on the stochastic vector bundles when the system performs work to reach its extremes. This is similar to the least action principle, a fundamental concept in classical mechanics. 

However, unlike energy conservation in classical Hamiltonian mechanics, stochastic systems consume energy to impose constraints on their evolution, quantified as the difference between the work done by the probability flux and the energy expended owing to the state transitions. The variation of \(\mathcal{S}\) implies that the system evolves to minimize the constraint represented by the energy residual \(\varepsilon=\wp_i\dot{y}^i-\mathrm{d}P(\dot{y})\), favoring paths that maximize changes in the probability distribution; essentially, paths requiring the least constraint. This result demonstrates the geometric nature underlying the evolution of stochastic systems, as described through the theory of contact manifolds and the associated contact dynamical equations.

\end{proof}

%
%
%
%
%
%

In fact, the study presented in this paper, especially Sections \ref{sec-3}, \ref{sec-4}, and what is included in this section about the discussion of \(\varepsilon\), leads to the following theorem:  
\\
\begin{theorem}[Contact Structure and Least Constraint for Stochastic Vector Bundles]\label{the-1}
Let \(E\) be an \(n\)-dimensional vector space and \(\mathbb{P}\) be the probability measure space. Subsequently, the infinite-order stochastic jet bundle \(\pi^{\infty}_{E,0}: J^{\infty}(E,\mathbb{P}) \rightarrow E\) exhibits the following properties:
\begin{enumerate}[label=(\roman*)]
    \item It has a natural contact manifold \((\mathcal{E}, \Theta)\) defined by the contact 1-form
    \[
    \Theta = \mathrm{d}P - \wp_i \mathrm{d}y^i = \mathcal{H} \mathrm{d}t - \wp_i \mathrm{d}y^i,
    \]
and the volume form \(\Theta \wedge (\mathrm{d}\Theta)^n\) is non-degenerate, characterizing a contact structure on \(J^{\infty}(E,\mathbb{P})\).
    \item There exists a vector field \(\mathfrak{X}(\mathcal{E}) = \{X_{\mathcal{H}} \mid X_{\mathcal{H}} \in \ker(\mathrm{d}\Theta) \subset T\mathcal{E}\}\) satisfying:
\[
\iota_{X_{\mathcal{H}}}\Theta = -\varepsilon, \quad
        \iota_{X_{\mathcal{H}}}\mathrm{d}\Theta = 0, 
\]
where \(\varepsilon\) is a smooth function termed as the constraint. The Lie derivative of \(\Theta\), i.e., \(L_{X_{\mathcal{H}}}\Theta = 0\) along the flow generated by \(\mathfrak{X}(\mathcal{E})\) preserves the contact structure.
    \item The flow generated by \(\mathfrak{X}(\mathcal{E})\) represents the extremal paths for the action of the constraint function \(\epsilon\),
    \[
    \mathcal{S} = \int \epsilon \mathrm{d}t = -\int \Theta,
    \]
indicating that the system evolves to minimize the constraint \(\epsilon\) while maximizing the variation in probabilities.
\end{enumerate}

\end{theorem}

\vspace{1em}
\begin{proof}
Proof of statement (i) is presented in Section \ref{sec-3} (Lemma \ref{lem-1}); 
proof of statement (ii) is outlined in Section \ref{sec-4} (Lemma \ref{lem-2}); 
and the proof of statement (iii) is incorporated in this section (Lemma \ref{lem-3}).
\\
\end{proof}

The least constraint theorem is a counterpart of the least action principle for symplectic structures: 

(1) The principle of least action arises from classical mechanics and is based on the concept of action, which is a measure of the ``effort" required to move a system from one state to another. By means of the variational principle, the actual path taken by a system between two states is the one that minimizes the action integral. This principle is formulated using Lagrangian or Hamiltonian formalism. It is widely used in classical mechanics to derive equations of motion for conservative systems such as planetary motion and harmonic oscillators. 

(2) In contrast, the theorem of least constraint is based on the concept of contact geometry. It is used to analyze the dynamics of stochastic systems in which energy conservation does not hold. It states that the evolution of a stochastic system tends to minimize the integration of a constraint function, which represents the residual energy or efficiency in the state transitions of the system. This theorem is developed using the contact geometry formalism, which describes the dynamics of the system in terms of the contact form as the constraint function.

Beyond its theoretical importance, the least constraint theorem is essential and applicable to various practical applications. For example, many deep learning algorithms involve agents that modify their behaviors or strategies within specified constraints to maximize changes in the probability distribution of their states or outputs, ultimately aiming to maximize cumulative rewards. This involves defining reward functions to identify actions that enhance the state transition probabilities. The least constraint theorem can be used to optimize policies, enabling the selection of strategies with the most significant environmental impact. Furthermore, it can balance exploration and exploitation to focus on actions that maximize probability changes or adjust strategies to respond dynamically to environmental shifts. In practical scenarios such as autonomous driving, deliberate actions can increase the likelihood of reaching a destination. This principle is expected to enhance the ability of these algorithms to make optimal decisions in complex environments, which is essential to improve the performance and adaptability of intelligent systems.
\section{Summaries and Conclusions}\label{sec-6}
This study investigates stochastic vector bundles, focusing on their infinite-order jet structure for geometric analysis and contact structures that support system evolution. It derives contact dynamical equations from the principle of least constraint, thereby providing a geometric framework for analyzing probabilistic systems across multiple fields.  

The following conclusions were drawn:

(1) This paper focuses on the contact structures of stochastic vector bundles, bridging differential geometry and stochastic processes. It aims to provide an approach to understanding stochastic processes through differential geometric principles, establishing connections between the geometric properties of vector bundles and the behavior of stochastic systems. The investigation examined contact structures using a 1-form, distinguishing them from symplectic and Riemannian structures. By integrating the theory of contact geometry, this study is expected to enrich research on stochastic processes. 

(2) This paper derives the contact dynamical equations for stochastic vector fields. They are intricately linked to the principle of least constraint, thereby elucidating the geometric structure underlying the evolution of stochastic systems and their tendency to minimize constraints.  Equations \eqref{eq-4-29a}, \eqref{eq-4-29b}, and \eqref{eq-4-29c} are essential for analyzing the dynamic behavior of stochastic processes. By estimating their probability distributions, these equations allow us to track system evolution over time, which is crucial in physics, statistics, and machine learning. This can also enhance our understanding of the underlying processes and aid physics-informed artificial intelligence (AI) modeling in diverse applications.

(3) The least constraint theorem is a counterpart of the least action principle for symplectic structure. The dynamical equations are intricately associated with the principle of least constraint, which elucidates the geometric structure underlying the evolution of stochastic systems and their tendency to minimize constraints. This tendency manifests as a more profound principle that governs the behavior of physical systems. This suggests that an underlying order and structure guide their evolution despite the inherent randomness and unpredictability of stochastic processes. By linking them to the theorem of least constraint, the contact dynamical equations provide a powerful tool for understanding this structure and predicting the behavior of stochastic systems under a wide range of conditions.

It is worth noting that the present study focuses on the theoretical aspects of the contact structure and dynamics of stochastic vector bundles. Therefore, only a simple example is presented in this study. Future investigations may involve exploring and developing robust numerical techniques for solving contact dynamical equations and applying them to study complicated dynamics of stochastic systems.

\bmhead{Acknowledgements} This study was supported financially by the State Key Laboratory of Hydroscience and Engineering of Tsinghua University under Grant No. sklhse-TD-2024-F01. We sincerely thank Mr. Mingxi Zhang and Ms. Ming Li for their invaluable assistance in preparing this paper. In particular, we sincerely appreciate the insightful comments the anonymous reviewers provided.

\section*{Declarations}

\begin{itemize}
\item \textbf{Conflict of interest}: The authors declare no Conflict of interest.
\item \textbf{Data availability}: Our manuscript has no associated data. 
\item \textbf{Publisher’s Note}: Springer Nature remains neutral with regard to jurisdictional claims in published maps and institutional affiliations. 

Springer Nature or its licensor (e.g., a society or other partner) holds exclusive rights to this article under a publishing agreement with the author(s) or other right-sholder(s); author self-archiving of the accepted manuscript version of this article is solely governed by the terms of such publishing agreement and applicable law.
\end{itemize}

%
%
%
\begin{appendices}
\section{Connection 1-form \(\wp\)}\label{app-1}
The concept of connection is significant in the theory of fiber bundles \cite{Nakahara2003GeometryTA, RUD2017}. It is a fundamental tool that enables us to differentiate vector fields along curves and understand how tensors change as we move from one point to another on a manifold. The connection allows us to define geodesics, which are curves that represent the shortest path between two points on a manifold. Geodesics generalizes the notion of straight lines in Euclidean space to curved spaces. In this study, the geodesic equation represents the curves in \(\mathrm{ker}(\Theta_\wp)\) owing to connection 1-form \(\Theta_{\wp}\), which determines the contact structure on stochastic vector bundles and is given as follows.

For the stochastic vector bundle of our study \cite{Zhong_2024}, on the one hand, \(\Theta_{\wp}\) is given by
\begin{equation}\label{eq-A-1}
\Theta_{\wp}=\mathrm{d}P_{\mu_1\cdots\mu_{k}}-\wp(P_{\mu_1\cdots\mu_{k}}, {y})
=
\mathrm{d}P_{\mu_1\cdots\mu_{k}}-\wp_i\mathrm{d}y^i, \: 0\le k\le \infty.
\end{equation}
Bearing in mind that \(\wp(P_{\mu_1\cdots\mu_{\infty}}, {y})\) is the connection 1-form of \(J^{\infty}(E,\mathbb{P})\), we had the geodesic equation for \(\mathbb{P}\) as:
\begin{equation}\label{eq-A-2}
	\frac{\mathrm{d}P(t, y(s))}{\mathrm{d}s}
	=\wp(\dot{y})=\wp_i\dot{y}^i.
\end{equation}
On the other hand, our study also showed that
\begin{equation}\label{eq-A-3}
\wp(y, P_{\mu_1\cdots\mu_{\infty}})=\pi^{*}\mathscr{L}P,
\end{equation}
which can be adopted to determine \(\wp\) in the connection 1-form. 

To obtain an explicit expression for \(\wp\) in this study, we re-write $P(t,y)$ in equation \eqref{eq-3-1} as a functional of a parameterized path $y=y(s)$: $P=P(t, y(s))$; correspondingly, 
\begin{equation}\label{eq-A-4}
	\frac{\partial P(t,y(s))}{\partial t}-\mathscr{L}(\dot{x}(s))P(t,y(s))=0,
\end{equation}
where \(s\) is the path parameter, and \(x=\pi_*y\). For a fixed time \(t\), differentiating equation \eqref{eq-A-4} with respect to the parameter $s$, $0\le s\le t$, and using the relation 
\begin{equation}\label{eq-A-5}
\frac{\mathrm{d} P(t, y(s))}{\mathrm{d} s}=\dot{y}^i(s)\frac{\partial P}{\partial y^i}=\mathrm{d}P(\dot{y}(s)),	
\end{equation}
we derived that 
\begin{equation}\label{eq-A-6}
\begin{split}
	\frac{\partial}{\partial t}\mathrm{d}P(\dot{y}(s))-\frac{\partial}{\partial s }(\mathscr{L}(\dot{x}(s))P(t,y(s)))=0.		
\end{split}
\end{equation}
Equation \eqref{eq-A-6} is also valid for $s=t$, and thus
\begin{equation}\label{eq-A-7}
\begin{split}
	\frac{\partial}{\partial t}\left[\mathrm{d}P(\dot{y}(t))-\mathscr{L}(\dot{x}(t))P(t,y(t))\right]=0.		
\end{split}
\end{equation}
Integration of equation \eqref{eq-A-7} results in
\begin{equation}\label{eq-A-8}
\begin{split}
	\mathrm{d}P(\dot{y}(t))=\mathscr{L}(\dot{x}(t))P(t,y(t))+C,		
\end{split}
\end{equation}
where \(C\) is the integration constant. For a dissipative system, since that \(t\to \infty\), \(\mathrm{d}P\to 0\) and \(\mathscr{L}P=0\), which means the probabilities are equal everywhere, we deduced that \(C=0\). 

If expressing \(\mathscr{L}\) of equation \eqref{eq-A-8} in terms of jet bundle notions (see equation \eqref{eq-2-3}), we found it can be written in an equivalent form as:  
\begin{equation}\label{eq-A-9}
\begin{split}
	\mathrm{d}P(\dot{y})
	=\sum_{0\le k\le \infty}\sum_{\mu_1<\cdots<\mu_k}D^{\mu_1\cdots\mu_ki}P_{\mu_1\cdots\mu_ki}(y),
\end{split}
\end{equation}
where 
\begin{equation}\label{eq-A-10}
\begin{split}
D^{\mu_1\cdots\mu_{k}i}
&=\frac{(-1)^{k+1}}{(k+1)!}\pi^{*}\left\langle\left\langle\underbrace{\int\cdots\int}_{\text{$k-$tuples}} \mathcal{L}(y^{\mu_1})\cdots\mathcal{L}(y^{\mu_k})\mathcal{L}(y^{i})\right\rangle\right\rangle.
\end{split}
\end{equation}
\(\mathcal{L}\) in equation \eqref{eq-A-10} is a vector-valued 1-form, and its pull-back by \(\pi^*\) is \cite{Zhong_2024}:  
\begin{equation}\label{eq-A-11}
\begin{split}
	\pi^{*}\mathcal{L}
	=\partial_i\otimes\tensor{\mathcal{L}}{^i_j}\mathrm{d}Y^j,	
\end{split}	
\end{equation}
where
\begin{equation}\label{eq-A-12}
\tensor{\mathcal{L}}{^i_j}=\tensor{\delta}{^i_j}\pi^{*}+\tensor{A}{^i_{\mu\nu}}Y^{\mu}\frac{\mathrm{d}X^{\nu}}{\mathrm{d}Y^j},	
\end{equation}
and within which \(\tensor{A}{^i_{\mu\nu}}\) is the coefficient of the local connection 1-form \(A\) for vector bundle \(\pi: E\to M\) \cite{Zhong_2024}; given a point \(p \in M\), \(X\in T_pM\), \(Y\in T_pE\), and \(\pi_*: Y\to X\). When it takes value at \(y^i\) and acts on \(\dot{y}\) as a 1-form, it is 
\begin{equation}\label{eq-A-13}
    \pi^{*}\mathcal{L}(y^i)(\dot{y})
	=\tensor{\mathcal{L}}{^i_j}\mathrm{d}Y^j(\dot{y})\partial_i(y^i).
\end{equation}
Therefore, using equation \eqref{eq-A-13}, we had that 
\begin{equation}\label{eq-A-14}
\begin{split}
	&\pi^{*}\left\langle\left\langle\underbrace{\int\cdots\int}_{\text{$k-$tuples}} \mathcal{L}(y^{\mu_1})\cdots\mathcal{L}(y^{\mu_k})\mathcal{L}(y^{i})\right\rangle\right\rangle \\
	=&
	\left\langle\left\langle\underbrace{\int\cdots\int}_{\text{$k-$tuples}} \left\{\pi^{*}[\mathcal{L}(y^{\mu_1})\cdots\mathcal{L}(y^{\mu_k})]\right\}\tensor{\mathcal{L}}{^i_j}\mathrm{d}Y^j\right\rangle\right\rangle  \\	
	=&
	\left\langle\left\langle\underbrace{\int\cdots\int}_{\text{$k-$tuples}} \left\{\pi^{*}[\mathcal{L}(y^{\mu_1})\cdots\mathcal{L}(y^{\mu_k})]\right\}\tensor{\mathcal{L}}{^i_j}\frac{\mathrm{d}Y^j}{\mathrm{d}y^l}\right\rangle\right\rangle \mathrm{d}y^l. 
\end{split}
\end{equation}
Let 
\begin{equation}\label{eq-A-15}
	{B}^{\mu_1\cdots\mu_kj}_i=\frac{(-1)^{k+1}}{(k+1)!}\left\langle\left\langle\underbrace{\int\cdots\int}_{\text{$k-$tuples}} \left\{\pi^{*}[\mathcal{L}(y^{\mu_1})\cdots\mathcal{L}(y^{\mu_k})]\right\}\tensor{\mathcal{L}}{^j_l}\frac{\mathrm{d}Y^l}{\mathrm{d}y^i}\right\rangle\right\rangle,
\end{equation}
then
\begin{equation}\label{eq-A-16}
\begin{split}
D^{\mu_1\cdots\mu_{k}i}
&=\frac{(-1)^{k+1}}{(k+1)!}\pi^{*}\left\langle\left\langle\underbrace{\int\cdots\int}_{\text{$k-$tuples}} \mathcal{L}(y^{\mu_1})\cdots\mathcal{L}(y^{\mu_k})\mathcal{L}(y^{i})\right\rangle\right\rangle\\
&={B}^{\mu_1\cdots\mu_kj}_i\mathrm{d}y^i,
\end{split}
\end{equation}
and thus 
\begin{equation}\label{eq-A-17}
\begin{split}
	\mathrm{d}P
	&=\sum_{0\le k\le \infty}\sum_{\mu_1<\cdots<\mu_k}D^{\mu_1\cdots\mu_ki}P_{\mu_1\cdots\mu_ki}(y)\\
	&=\sum_{0\le k\le \infty}\sum_{\mu_1<\cdots<\mu_k}{B}^{\mu_1\cdots\mu_kj}_iP_{\mu_1\cdots\mu_kj}\mathrm{d}y^i\\
	&=\sum_{1\le k\le \infty}\sum_{\mu_1<\cdots<\mu_k}{B}^{\mu_1\cdots\mu_k}_iP_{\mu_1\cdots\mu_k}\mathrm{d}y^i.
\end{split}
\end{equation}
Finally, comparison of \eqref{eq-A-2} and \eqref{eq-A-17} results in
\begin{equation}\label{eq-A-18}
\wp=\sum_{1\le k\le \infty}\sum_{\mu_1<\cdots<\mu_k}{B}^{\mu_1\cdots\mu_k}_iP_{\mu_1\cdots\mu_k}\mathrm{d}y^i,
\end{equation}
which is the local connection 1-form for \(\pi^{\infty}_{E,0}: J^{\infty}(\mathbb{P}, E)\to E\), as we expected that it is a linear combination of \(P_{\mu_1\cdots\mu_k}\).
\end{appendices}
\bibliography{bib-Contact-2024}
\end{document}